\documentclass[twocolumn,showpacs,preprintnumbers,amsmath,amssymb]{revtex4}

\usepackage{graphicx}
\usepackage{dcolumn}
\usepackage{bm}

\begin{document}

\preprint{APS/123-QED}

\title{Yb$_{3}$Pt$_{4}$: A New Route to Quantum Criticality}
\author{M. C. Bennett,$^{1,2}$ D. A. Sokolov,$^{1,3}$ M. S. Kim,$^{1,3}$ Y. Janssen,$^{3}$ Yuen Yiu,$^{1,2}$ W. J. Gannon,$^{1}$ and M. C. Aronson$^{1,2,3}$}

\affiliation{$^{1}$ Department of Physics, University of Michigan, Ann Arbor, MI 48109-1120}
\affiliation{$^{2}$ Department of Physics and Astronomy, Stony Brook University, Stony Brook, NY 11974}
\affiliation{$^{3}$ Brookhaven National Laboratory, Upton, NY  11973}

\date{\today}

\begin{abstract}
We have studied the evolution of the weakly first order
antiferromagnetic transition in heavy fermion Yb$_{3}$Pt$_{4}$ using
a combination of specific heat, magnetic susceptibility, and
electrical resistivity experiments. We show that magnetic fields
suppress the Neel temperature, as well as the specific heat jump,
the latent heat, and the entropy of the transition, driving a
critical endpoint at 1.2 K and 1.5 T. At higher fields, the
antiferromagnetic transition becomes second order, and this line of
transitions in turn terminates at a quantum critical point at 1.62
T.  Both the ordered and high field paramagnetic states are Fermi
liquids at low temperature, although the former has a much larger
magnetic susceptibility and stronger quasiparticle scattering.
Unlike previously studied quantum critical systems, the
quasiparticle mass in Yb$_{3}$Pt$_{4}$ does not diverge at the
quantum critical point, implying that here the quasiparticle
interactions drive the zero temperature transition. The Fermi liquid
parameters are nearly field-independent in the ordered state,
indicating that the fluctuations never become fully critical, as
their divergences are interrupted by the first order
antiferromagnetic transition.
\end{abstract}

\pacs{75.30.Mb, 75.20.Hr, 71.27.+a}
\maketitle

Quantum critical points (QCPs) are increasingly
recognized~\cite{coleman2005} as organizing features of the phase
diagram of strongly interacting electronic systems, from heavy
electrons,~\cite{stewart2001,vonlohneysen2007,gegenwart2008} to
complex oxides,~\cite{laughlin1998,varma1989} and low dimensional
conductors.~\cite{jaccard2001} Since the QCP is a T=0 phase
transition driven not by reduced temperature, but by varying
parameters such as pressure or magnetic field, it is heralded by
unusual non-Fermi liquid temperature dependences in measured
quantities such as the electrical resistivity $\rho$, magnetic
susceptibility $\chi$, and specific heat
C~\cite{stewart2001,vonlohneysen2007} which derive from quantum
critical phenomena.~\cite{millis1993}It is believed that the
associated quantum critical fluctuations are crucial for stabilizing
unconventional electronic states near QCPs, such as magnetically
mediated superconductivity~\cite{mathur1998,saxena2000} and electron
nematic phases~\cite{borzi2007}.

So far, magnetic phase diagrams hosting QCPs are found to fall into
two distinct classes. Itinerant ferromagnets have been carefully
studied, and in the cleanest systems the very lowest temperature
phase transitions are first order,~\cite{pfleiderer2005,uemura2007}
in agreement with theoretical expectations.~\cite{belitz2005} The
implication is that the fluctuations near the quantum phase
transition may not actually be fully critical, i.e. infinite in
range and lifetime. In contrast, there are many examples among the
heavy electron compounds of quantum critical antiferromagnets, where
the Neel temperature can be  driven to zero by an external parameter
such as pressure or magnetic
field.~\cite{julian1996,vonlohneysen1996,gegenwart2003a,budko2004}The
success here of scaling analyses, particularly of neutron scattering
data,~\cite{aronson1995,aronson2001,schroder2000} provides direct
evidence for the dominance of quantum critical fluctuations in these
compounds. While it is possible that closer scrutiny in cleaner
systems will reveal as-yet hidden first order character, at present
there are no known exceptions to the current theoretical and
experimental consensus that all T=0 antiferromagnets are second
order, i.e. genuinely quantum critical.

The experimental record is clear that in these quantum critical
antiferromagnets, the magnetic state emerges at T=0 from a Fermi
liquid state, where the quasiparticle mass diverges as the tuning
variable approaches its critical
value~\cite{steglich1997,gegenwart2003b,nakamura2006,coleman2001}.
In some cases, this implied electronic localization is reflected in
a discontinuous change in the Fermi surface volume at the
QCP.~\cite{shishido2005,paschen2004} We stress that much of this
understanding has been gleaned from experimental and theoretical
studies of antiferromagnets which are truly quantum critical, and
that a very different development of the electronic structure may be
possible in antiferromagnets where fluctuations play a less decisive
role. Careful studies of the behaviors found in different parts of
the phase diagram generated by tuning a first order
antiferromagnetic transition are consequently very important, but
only recently has a suitable experimental system been discovered.

We have recently reported that Yb$_{3}$Pt$_{4}$ undergoes a weakly
first order transition into an antiferromagnetic state with a Neel
temperature T$_{N}$=2.4 K, which is also a Fermi
liquid.~\cite{bennett2008} Magnetic order develops from a simple
paramagnetic state where the magnetic susceptibility obeys a Curie
law, indicating that the Yb f-electrons are excluded from the Fermi
surface for T$\geq$T$_{N}$. The crystal field scheme deduced from
specific heat measurements reveals a well isolated magnetic doublet
ground state, and the entropy associated with magnetic order is
$\sim$0.8 Rln2. We present here detailed measurements of the
specific heat C, ac and dc magnetic susceptibility $\chi$, and
electrical resistivity $\rho$ using magnetic fields to tune the
stability of antiferromagnetic order. The phase diagram revealed by
these measurements is very different from any reported previously,
with a line of first order transitions terminating at a critical
endpoint, from which a line of second order transitions continues to
a QCP. At low temperatures, both the antiferromagnet and paramagnet
are Fermi liquids, which evolve very differently as the QCP is
approached.

\begin{figure}[t]
\begin{center}
\includegraphics[width=9.8cm]{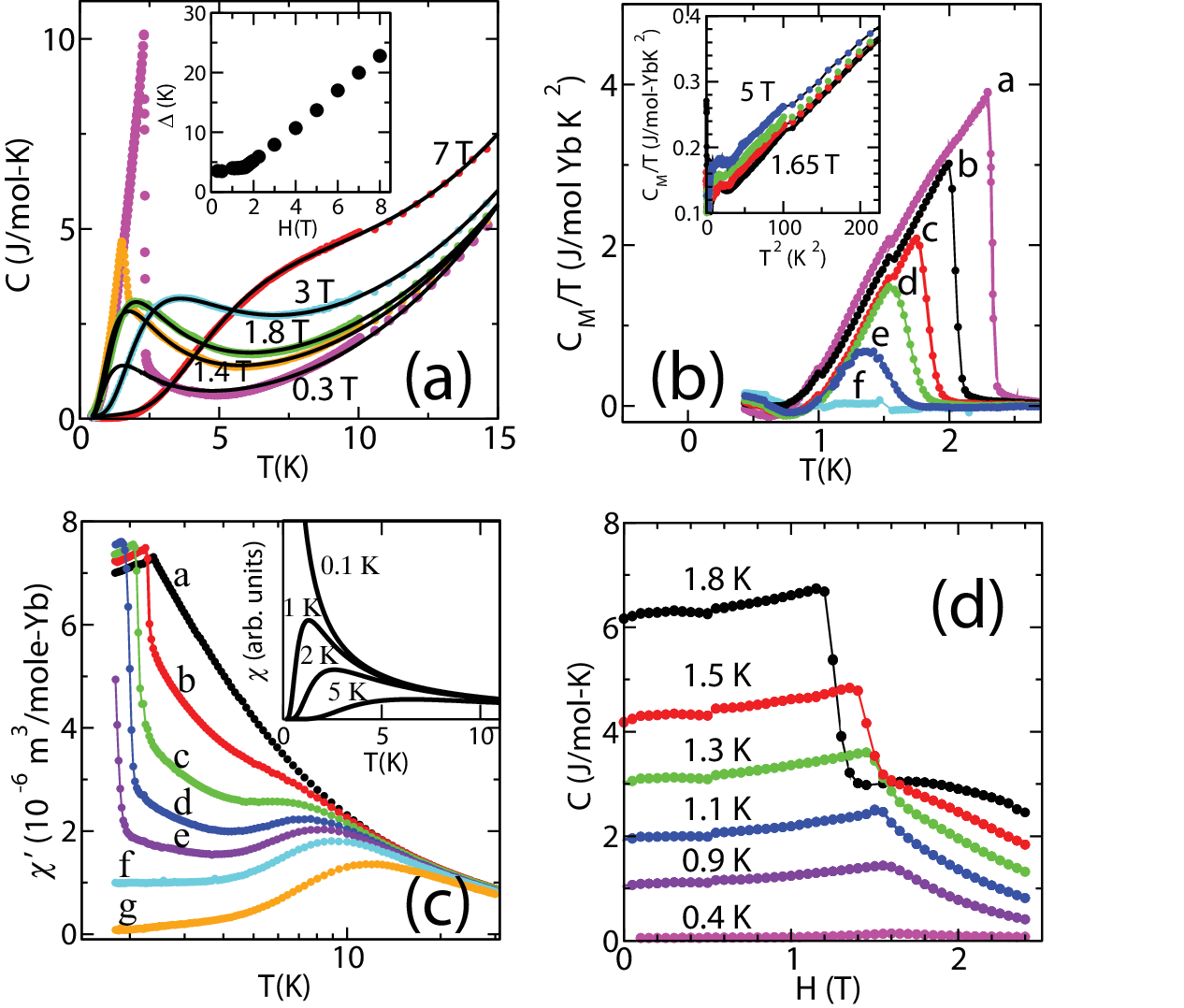}
\end{center}
\caption{\label{f1} (a) The specific heat C of Yb$_{3}$Pt$_{4}$
measured in magnetic fields H$\parallel$a. Solid lines are fits to
C(T)=C$_{Sch}$+$\gamma$T+$\beta$T$^{3}$. Inset: The splitting
$\Delta$ of the ground state doublet in field. (b)
C$_{M}$/T=(C-C$_{Schottky}$)/T in different fields(a=0.3 T, b=1 T,
c=1.25 T, d=1.37 T, e=1.5 T, f=1.65 T). Inset: C$_{M}$/T plotted as
a function of T$^{2}$ for H=1.65 T, 2.25 T, 3 T, and 5 T. The steps
near 10 K reflect a systematic error in the PPMS temperature
calibrations.(c) The real part of the ac susceptibility $\chi\prime$
for different fields H $\parallel$a (a=0.1 T, b=0.6 T,c=0.9 T,d=1.1
T, e=1.3 T, f=1.6 T, g=2.5 T). Inset: Calculated $\chi$(T) for
different values of $\Delta$, the splitting of an isolated Yb$^{3+}$
doublet.(d) The field dependence of the specific heat C
(H$\parallel$a) at fixed temperatures 0.4 K - 1.8 K.}
\end{figure}

\begin{figure}[t]
\begin{center}
\includegraphics[width=9.5cm]{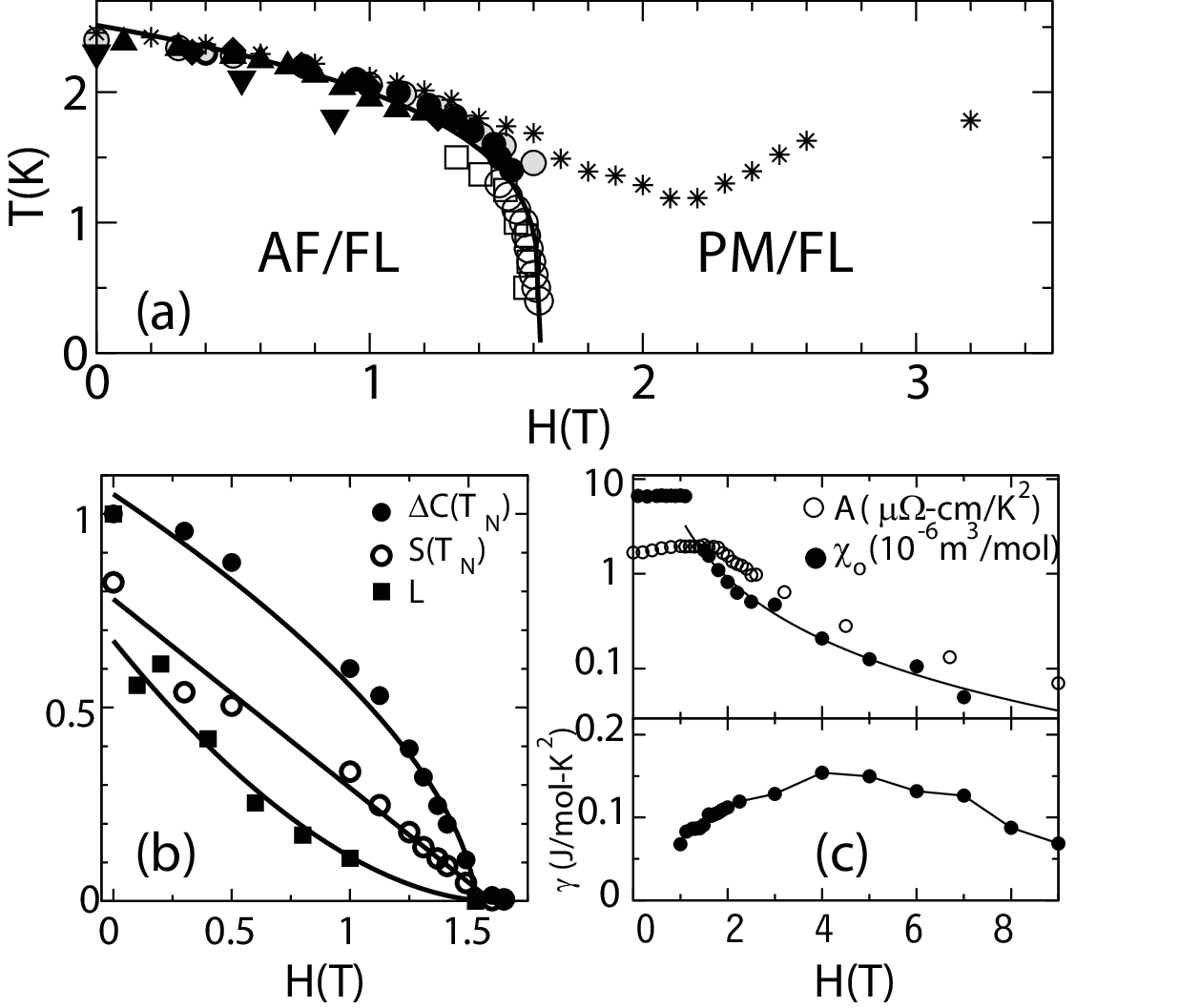}
\end{center}
\caption{\label{f2}(a) The magnetic phase diagram of
Yb$_{3}$Pt$_{4}$ for H$\parallel$a taken from temperature sweeps of
$\chi$($\blacktriangle$), dM/dT($\blacklozenge$), C(T)(gray circles)
and field sweeps of C(H)(1$^{st}$ order: $\bullet$, 2$^{d}$ order:
$\bigcirc$) and $\rho$(H)(1$^{st}$ order: $\blacktriangledown$,
2$^{d}$ order: $\Box$). T$_{FL}$ ($\ast$) is the limit of
$\Delta\rho\sim$T$^{2}$. The first and second order lines are
jointly fit by T$_{N} \sim$(H-1.62 T)$^{0.28}$ (solid line).
(b)$\Delta$C(T$_{N}$)(normalized to zero field value 8.82 J/mol-K,
filled circles), entropy S(T$_{N}$) (units of Rln2, open circles),
latent heat L (normalized to zero field value 0.09 J/mol-Yb, filled
squares) Lines are guides for the eye. (c) Field dependencies of the
Fermi liquid parameters A, $\chi_{0}$, and $\gamma$, defined in the
text. Solid line for $\chi_{0} \propto 1/H^{3}$, solid line for
$\gamma$ is guide for eye.}
\end{figure}

Measurements of the specific heat C provide the first indication
that magnetic fields oriented along the a-axis tune criticality in
Yb$_{3}$Pt$_{4}$. Fig.\ref{f1}a shows that at every field C is well
described by the sum of an electronic term $\gamma$T, a field
independent phonon contribution $\beta$T$^{3}$, a broad Schottky
contribution C$_{Sch}$, and a residual magnetic and electronic
contribution C$_{M}$ (Fig.\ref{f1}b), in which the onset of
antiferromagnetic order at T$_{N}$ is marked by a step-like anomaly
$\Delta$C(T$_{N}$). Magnetic fields reduce T$_{N}$ (Fig.\ref{f2}a),
while $\Delta$C and the latent heat L of the transition are
suppressed continuously to zero at 1.5 T(Fig.\ref{f2}b). The broad
Schottky peak traverses our temperature window with increasing
field, and the inset of Fig.\ref{f1}a shows the related Zeeman
splitting $\Delta$ of the ground doublet. We have plotted the
entropy S associated with C$_{M}$ in Fig.\ref{f2}b, which is reduced
from its zero field value of $\sim$ 0.8 Rln2 to zero in a field of
1.5 T. Taken together, these data show that all thermodynamic
signatures of the first order transition vanish simultaneously at
1.5 T, although T$_{N}$ remains nonzero. We conclude that the line
of first order antiferromagnetic transitions which originates from
the zero field value of 2.4 K extends to a critical endpoint at 1.2
K and 1.5 T (H$\parallel$ a). Similar measurements reveal that the
critical endpoint occurs at 1.2 K and 3.5 T when H$\parallel$c.


The first order phase line T$_{N}$(H) is also evident in other
measurements. While the magnetization M is continuous at T$_{N}$(H),
its field and temperature derivatives are discontinuous. The
temperature dependence of the real part of the ac susceptibility
$\chi\prime$ is plotted for different fixed fields in Fig.\ref{f1}c.
In low fields, an antiferromagnetic cusp is observed at the Neel
temperature T$_{N}$. With increased field, steplike discontinuities
in $\chi\prime$ and also in dM/dT (not shown here) occur at the same
ordering temperatures found in C (Fig.\ref{f2}a), in both cases
becoming stronger as the field approaches the critical end point at
1.5 T. Fig.\ref{f1}c shows that these manifestations of the phase
transition in $\chi\prime$ are superposed on a field and temperature
dependent background, which is Curie-like at low fields, and then
develops a broad maximum which moves to higher temperatures with
increasing field. In every field, Curie-law behavior is regained at
high temperatures. The inset of Fig.\ref{f1}c shows a calculation of
$\chi$(T), assuming a two level system with different values of the
doublet splitting $\Delta$ which qualitatively reproduces this
behavior.

\begin{figure}[t]
\begin{center}
\includegraphics[width=8.75cm]{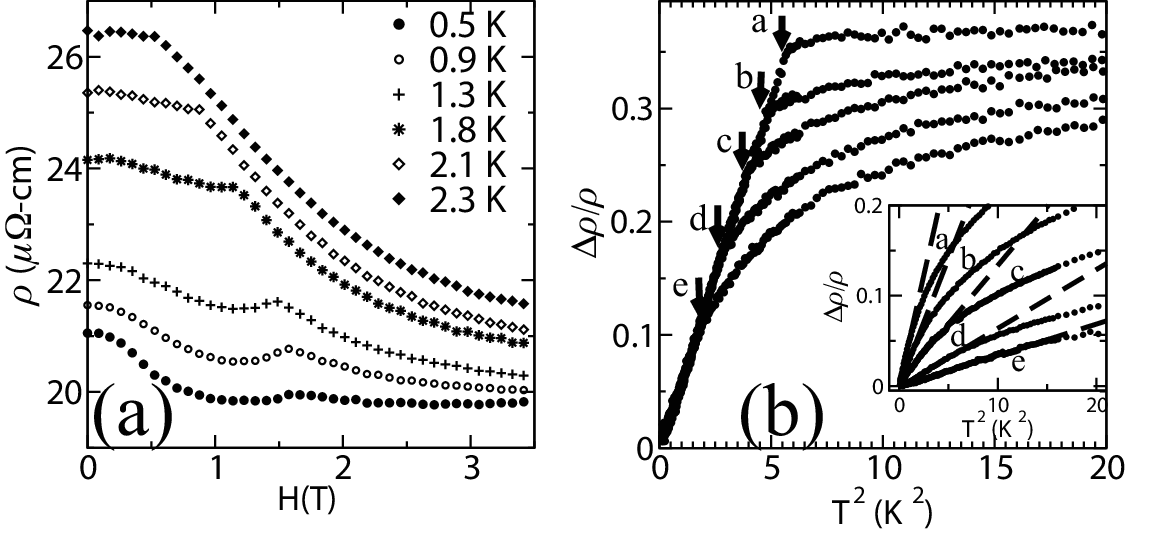}
\end{center}
\caption{\label{f3}(a) The field dependence of the electrical
resistivity $\rho$ for fixed temperatures 0.5 K - 2.3 K
(H$\parallel$a).
(b)$\Delta\rho$/$\rho$=\mbox{$\rho$(T)-$\rho_{0}$/$\rho_{0}$} is
linear in T$^{2}$ up to T$_{FL}$=T$_{N}$(marked by arrows).
(a,b,c,d,e correspond to 0, 0.8T, 1.2 T, 1.5 T, 1.8 T). Inset: same,
but with H$\geq$H$_{QCP}$=1.62 T.(a,b,c,d,e correspond to 2.3 T, 3.2
T, 4.5 T, 6.7 T, 9T). Dashed lines are linear fits.}
\end{figure}

It is of great interest to determine whether the first order phase
line extends beyond the critical endpoint as one or more lines of
second order phase transitions, or simply gives way to a coexistence
regime. Accordingly, C(H) is plotted at different fixed temperatures
in Fig. \ref{f1}d. For temperatures which are larger than the 1.2 K
critical endpoint(T$_{CEP}$), a step is observed in C(H),
reminiscent of the step in C(T) shown in Fig.\ref{f1}b. For
T$\leq$T$_{CEP}$, the step becomes a lambda-like anomaly, as
expected for a second order phase transition. This peak in C(H)
moves to slightly higher fields and is reduced in amplitude as the
temperature is reduced. The peaks in C(H) shown in Fig.\ref{f1}d
trace out a line of second order transitions which terminates at a
zero temperature QCP with a critical field H$_{QCP}$ of $\sim$1.6T
(Fig.\ref{f2}a). Further evidence for a second order line emanating
from the critical endpoint comes from the field dependence of the
electrical resistivity $\rho$(H), shown Fig.\ref{f3}a. For scans
across the first order phase line (T$\geq$1.2 K), $\rho$ is
continuous, like M, but d$\rho$/dH is discontinuous, like
$\chi$=dM/dH. For lower temperatures (T$\leq$1.2 K) where we are
scanning across the second order phase line, $\rho$(H) develops a
cusp, signalling an incipient divergence of the magnetization. Taken
together, the first and second order lines are well described by the
expression T$_{N} \propto$ (H-H$_{QCP}$)$^{0.28\pm 0.03}$ with
H$_{QCP}$=1.62 T. We note that this critical exponent is much
smaller than 2/3, the value expected for a mean field quantum
critical antiferromagnet,~\cite{millis1993} and does not reproduce
the quasi-linear suppression of T$_{N}$ found in other quantum
critical antiferromagnets.~\cite{vonlohneysen1996, julian1996}

Our measurements indicate that both the antiferromagnetic state and
the high field paramagnetic state are Fermi liquids. We have tracked
their development by measuring the temperature dependence of
$\Delta\rho$/$\rho$=($\rho$(T)-$\rho$(T=0))/$\rho$(T=0) in fixed
fields both above and below the quantum critical field
H$_{QCP}$=1.62 T (Fig.\ref{f3}b). $\Delta\rho$=A(H)T$^{2}$ below an
effective Fermi temperature T$_{FL}$ which is indistinguishable from
T$_{N}$ for fields H$\leq$H$_{QCP}$ (Fig. \ref{f2}a). Long lived and
long ranged critical fluctuations generally suppress Fermi liquid
behavior near the onset of order, so the resilience of the
antiferromagnetic Fermi liquid in Yb$_{3}$Pt$_{4}$ is remarkable,
perhaps reflecting a general weakness of critical fluctuations near
a first order transition. The quadratic temperature dependence for
$\Delta\rho$ extends well into the paramagnetic phase
H$\geq$H$_{QCP}$, with T$_{FL}$ reaching its minimum value near the
critical endpoint before increasing again with increasing magnetic
fields. Fig.\ref{f3}b also shows that the resistivity coefficient
A(H) is almost field independent within the antiferromagnetic phase
H$\leq$H$_{QCP}$, and then drops rapidly at larger fields. These
observations are quantified in Fig. \ref{f2}c, where A decreases by
a factor of $\sim$ 30 from its value in the antiferromagnetic low
field phase to its value in 9 T.

The magnetic susceptibility also suggests that the ordered state is
a Fermi liquid. Fig.\ref{f1}c shows that $\chi\prime$ for
T$\leq$T$_{N}$ consists of a large and temperature independent term
$\chi_{0}$ and a much weaker temperature dependent contribution. No
trace is found in the antiferrromagnetic state of the local moment
behavior found in the susceptibility for T$\geq$T$_{N}$, suggesting
that here the f-electron has been absorbed into the Fermi surface.
$\chi_{0}$ is plotted as a function of field in Fig.\ref{f2}c. While
slight variations in field alignment are likely responsible for the
slightly different values of H$_{QCP}$ found in the different
measurements, $\chi_{0}$, like A, is nearly field independent in the
ordered state, but falls off dramatically in the paramagnetic state
(H$\geq$H$_{QCP}$).

The interpretation of the specific heat data is more complex. The
convolution of the Schottky and ordering peaks makes it impossible
to unambiguously isolate the Fermi liquid part of the specific heat
at the lowest fields. We limit our analysis to fields larger than
1.4 T, where the inset to Fig.\ref{f1}b confirms there is a linear
contribution to C with a magnitude $\gamma$ which increases from
$\sim$60 mJ/mol-K$^{2}$ at 1.4 T to a maximum value of 160
mJ/mol-K$^{2}$ at 5 T, before falling off at higher fields
(Fig.\ref{f2}c).

\begin{figure}[t]
\begin{center}
\includegraphics[width=5.5cm]{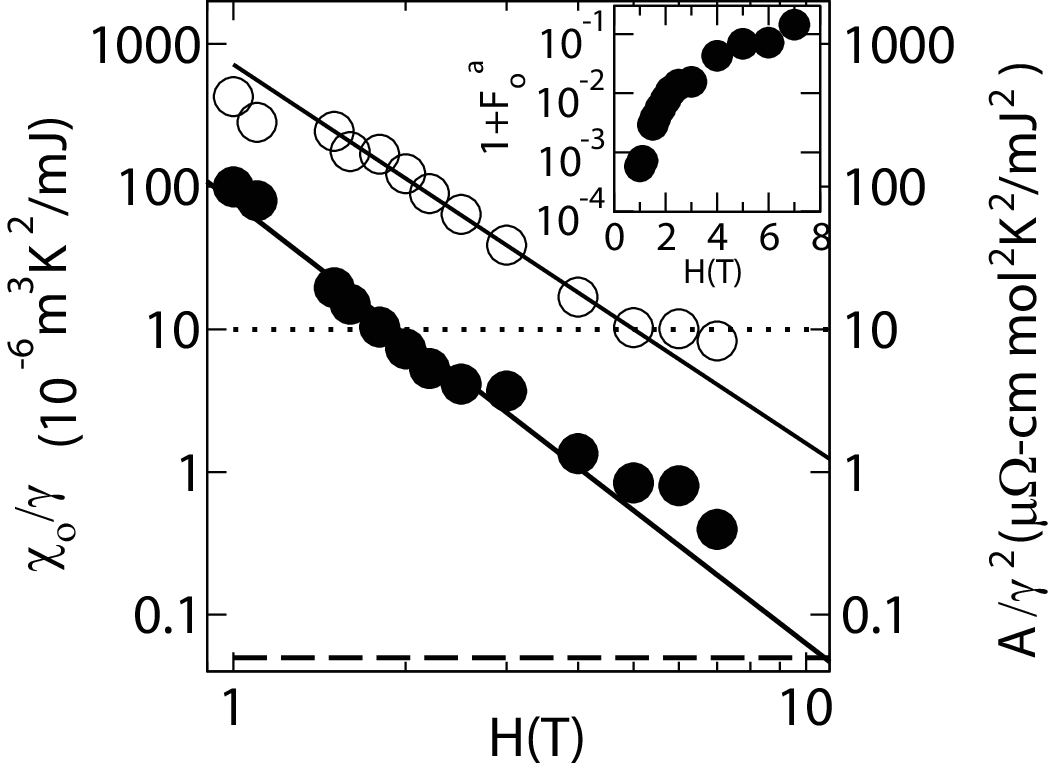}
\end{center}
\caption{\label{f4} The field dependencies of the Sommerfeld-Wilson
ratio $\chi_{0}$/$\gamma$ (filled circles, left axis) and the
Kadowaki-Woods ratio A/$\gamma^{2}$ (open circles, right axis).
Solid lines are power law fits, described in the text. Dotted line:
A/$\gamma^{2}$=10$\mu\Omega$-cm mol$^{2}$K$^{2}$/J$^{2}$), dashed
line: $\chi_{0}$/$\gamma$=5x10$^{-8}$m$^{3}$K$^{2}$/mJ. Inset: The
field dependence of 1+F$_{0}^{a}$.}
\end{figure}
The field evolution of the Fermi liquid parameters A, $\chi_{0}$ ,
and $\gamma$ reveals an unusual sequence of events which accompany
the onset of antiferromagnetic order in Yb$_{3}$Pt$_{4}$. Unlike
other systems where the destruction of the quasiparticles at the QCP
is mirrored in the divergence of the quasiparticle mass, in
Yb$_{3}$Pt$_{4}$ $\gamma$ is only weakly field dependent. Since
$\chi_{0}\propto$ \mbox{m$^{\ast}$/m (1/1+F$_{0}^{a}$)}, we conclude
that in the absence of a diverging quasiparticle mass enhancement
m$^{\ast}$/m, the field dependence of $\chi_{0}$ must be dominated
in Yb$_{3}$Pt$_{4}$ by that of (1/1+F$_{0}^{a}$). Fig. 4 shows that
$\chi_{0}$/$\gamma$ increases from a value of 5x10$^{-8}$
m$^{3}$K$^{2}$/mJ at 9 T, typical of other heavy electron compounds
\cite{fisk1987}, to its final value of 2x10$^{-4}$ m$^{3}$K$^{2}$/mJ
at H$_{QCP}$, corresponding to a divergence
$\chi_{0}$/$\gamma\propto$1/(1+F$_{0}^{a}$)$\sim$H$^{-3\pm0.2}$. We
stress that this field dependence is not well described by
1/(H-H$^{\ast}$)$^{n}$ for any nonzero H$^{\ast}$, or for any n.
1/1+F$_{0}^{a}$ can be isolated using
R$_{W}$=$\pi^{2}$k$_{B}^{2}$$\chi_{0}$/($\mu_{0}\mu_{eff}\gamma$)=1/(1+F$_{0}^{a}$)
(Fig. \ref{f4},inset). 1+F$_{0}^{a}$ approaches zero, implying that
enhanced long-range interactions among the quasiparticles may drive
the T=0 phase transition, much as in a Stoner ferromagnet, or in
$^{3}$He itself. A matching divergence in the Kadowaki-Woods ratio
A/$\gamma^{2}\propto 1/H^{2.6\pm0.2}$ (Fig. \ref{f4}) reveals that
the normal heavy electron behavior A/$\gamma^{2}$=10$\mu\Omega$-cm
mol$^{2}$K$^{2}$/J$^{2}$ found at large fields~\cite{tsujii2003} is
supplanted by a sixty-fold increase in the quasiparticle -
quasiparticle scattering. If the diverging exchange field were
primarily ferromagnetic, as implied by F$_{o}^{a}\rightarrow$ -1
(Fig.4,inset), we would expect forward scattering to dominate the
transport, with A(H)$\propto\chi^{2}$.~\cite{dy1969,zwicknagl1992}
Instead we have A(H)$\propto \chi$, suggesting that quasiparticle
interactions with nonzero wave vector are likely also strengthening
on the approach to the antiferromagnetic state. Finally, the
saturation of A and $\chi_{0}$ in the ordered state with
H$\leq$H$_{QCP}$ signals that the first order onset of long range
antiferromagnetic order cuts off further development of these
quasiparticle correlations.

In conclusion, our data indicate that Yb$_{3}$Pt$_{4}$ is unique
among quantum critical systems studied so far. It has an unusual
phase diagram, where a weakly first order phase line terminates at a
critical endpoint, and is continued along a second order phase line
to a QCP, where the critical phenomena are emphatically not
mean-field like. The zero temperature magnetic susceptibility
increases rapidly as the field approaches this quantum critical
field, but this divergence is controlled by a fixed point at zero
field which is prematurely cut off by magnetic order, and no
evidence is found for strong fluctuations associated with the QCP.
Fermi liquid behavior is found on both sides of this field driven
QCP, although the antiferromagnetically ordered Fermi liquid at low
fields is more strongly interacting than the paramagnetic Fermi
liquid above the quantum critical field. Unlike previously studied
systems, there is no divergence of the quasiparticle mass at the
quantum critical field, although the interactions among the
quasiparticles strengthen dramatically, suggesting that they
interact with each other in a medium which becomes progressively
more susceptible, eventually culminating in magnetic order. This
scenario is unique among the heavy electron compounds studied so
far, and demonstrates that Yb$_{3}$Pt$_{4}$ represents a new route
to quantum criticality.

The authors acknowledge useful conversations with C. Varma, Q. Si,
P. Coleman, E. Abrahams, G. Zwicknagl, and V. Zlatic. Work at Stony
Brook University and the University of Michigan was supported by the
National Science Foundation under grant NSF-DMR-0405961.

\end{document}